\documentclass{article}
\usepackage{ismir,amsmath,cite,url}
\usepackage{amssymb}
\usepackage{graphicx}
\usepackage{color}

\usepackage{booktabs} 
\usepackage[group-separator={,}]{siunitx} 
\usepackage[flushleft]{threeparttable} 
\usepackage{tabularx}
\usepackage{enumitem}
\setdescription{font=\itshape}

\graphicspath{{figs/}}
\title{Evaluating language models of tonal harmony}

\multauthor
{David R. W. Sears$^1$ \hspace{1cm} Filip Korzeniowski$^2$ \hspace{1cm} Gerhard Widmer$^2$} { \\
	$^1$ College of Visual \& Performing Arts, Texas Tech University, Lubbock, USA \\
	$^2$ Institute of Computational Perception, Johannes Kepler University, Linz, Austria \\
	{\tt\small david.sears@ttu.edu}
}
\def\authorname{Sears, Korzeniowski, Widmer}

\begin{document}

\maketitle
\begin{abstract}
	This study borrows and extends probabilistic language models from natural language processing to discover the syntactic properties of tonal harmony. Language models come in many shapes and sizes, but their central purpose is always the same: to predict the next event in a sequence of letters, words, notes, or chords. However, few studies employing such models have evaluated the most state-of-the-art architectures using a large-scale corpus of Western tonal music, instead preferring to use relatively small datasets containing chord annotations from contemporary genres like jazz, pop, and rock.
	
	Using symbolic representations of prominent instrumental genres from the common-practice period, this study applies a flexible, data-driven encoding scheme to (1) evaluate Finite Context (or \textit{n}-gram) models and Recurrent Neural Networks (RNNs) in a chord prediction task; (2) compare predictive accuracy from the best-performing models for chord onsets from each of the selected datasets; and (3) explain differences between the two model architectures in a regression analysis. We find that Finite Context models using the Prediction by Partial Match (PPM) algorithm outperform RNNs, particularly for the piano datasets, with the regression model suggesting that RNNs struggle with particularly rare chord types.
\end{abstract}
\section{Introduction}\label{sec:introduction}

\begin{figure}[t!]
	\hspace{.1em}
	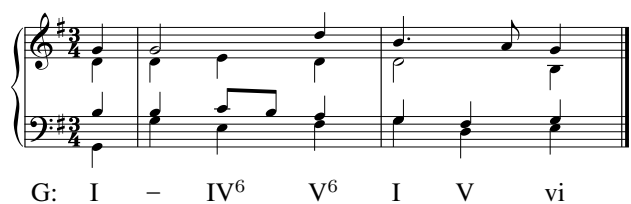
	
	\vspace{-.1cm}
	
	\caption{Bach, ``Aus meines Herzens Grunde'', mm. 1--2; from the Riemenschneider edition, No. 1. Key and Roman numeral annotations appear below.}
	\vspace{-.4cm}
	\label{fig:bach_example}
\end{figure}

For over two centuries, scholars have observed that tonal harmony, like language, is characterized by the logical ordering of successive events, what has commonly been called \textit{harmonic syntax}. 
In Western music of the common-practice period (1700-1900), pitch events group (or cohere) into discrete, primarily tertian sonorities, and the succession of these sonorities over time produces meaningful syntactic progressions. To characterize the passage from the first two measures of Bach's ``Aus meines Herzens Grunde'', for example, theorists and composers developed a chord typology that specifies both the scale steps on which tertian sonorities are built (\textit{Stufentheorie}), and the functional (i.e., temporal) relations that bind them (\textit{Funktionstheorie}). Shown beneath the staff in \figref{fig:bach_example}, this \textit{Roman numeral} system allows the analyst to recognize and describe these relations using a simple lexicon of symbols.

In the presence of such language-like design features, music scholars have increasingly turned to string-based methods from the natural language processing (NLP) community for the purposes of pattern discovery \cite{Conklin:2002}, classification \cite{Conklin:2013}, similarity estimation \cite{Mullensiefen:2009}, and prediction \cite{Pearce:2005}. In sequential prediction tasks, for example, probabilistic language models have been developed to predict the next event in a  sequence --- whether it consists of letters, words, DNA sequences, or in our case, chords.

Although corpus studies of tonal harmony have become increasingly commonplace in the music research community, applications of language models for chord prediction remain somewhat rare. This is likely because language models take as their starting point a sequence of chords, but the musical surface is often a dense web of chordal and nonchordal tones, making automatic harmonic analysis a tremendous challenge. Indeed, such is the scope of the computational problem that a number of researchers have instead elected to start with a particular chord typology right from the outset (e.g., Roman numerals, figured bass nomenclature, or pop chord symbols), and then identify chord events using either human annotators \cite{burgoyne_expert_2011}, or rule-based computational classifiers \cite{Temperley:1999}. As a consequence, language models for tonal harmony frequently train on relatively small, heavily curated datasets ($<200,000$ chords) \cite{burgoyne_expert_2011}, or use data augmentation methods to increase the size of the corpus \cite{Korzeniowksi:2018}. And since the majority of these corpora reflect pop, rock, or jazz idioms, vocabulary reduction is a frequent preliminary step to ensure improved model performance, with the researcher typically including specific chord types (e.g., major, minor, seventh, etc.), thus ignoring properties of tonal harmony relating to inversion \cite{Korzeniowksi:2018} or chordal extension \cite{DiGiorgi:2017}.


Given the state of the annotation bottleneck, we propose a complementary method for the implementation and evaluation of language models for chord prediction. Rather than assume a particular chord typology a priori and train our models on the \textit{chord classes} found therein, we will instead propose a data-driven method for the construction of harmonic corpora using \textit{chord onsets} derived from the musical surface. It is our hope that such a bottom-up approach to chord prediction could provide a springboard for the implementation of chord class models in future studies \cite{Brown:1992a}, the central purpose of which is to use predictive methods to reduce the musical surface to a sequence of syntactic progressions by discovering a small vocabulary of chord types. 

We begin in Section \ref{sec:corpus} by describing the datasets used in the present research and then present the tonal encoding scheme that reduces the combinatoric explosion of potential chord types to a vocabulary consisting of roughly two hundred types for each scale-degree in the lowest instrumental part. Next, Section \ref{sec:languagemodels} describes the two most state-of-the-art architectures employed in the NLP community: Finite Context (or \textit{n}-gram) models and Recurrent Neural Networks (RNNs). Section \ref{sec:experiments} presents the experiments, which (1) evaluate the two aforementioned model architectures in a chord prediction task; (2) compare predictive accuracy from the best-performing models for each dataset; (3) attempt to explain the differences between the two models using a regression analysis. We conclude in Section \ref{sec:discussion} by considering limitations of the present approach, and offering avenues for future research.


\section{Corpus}\label{sec:corpus}

This section presents the datasets used in the present research and then describes the chord representation scheme that permits model comparison across datasets.

\subsection{Datasets}

Shown in \tabref{tab:corpus}, this study includes nine datasets of Western tonal music (1710--1910) featuring symbolic representations of the notated score (e.g., metric position, rhythmic duration, pitch, etc.). The Chopin dataset consists of 155 works for piano that were encoded in MusicXML format \cite{Flossmann:2010b}. The Assorted symphonies dataset consists of symphonic movements by Beethoven, Berlioz, Bruckner, and Mahler that were encoded in MATCH format \cite{Widmer:2001}. All other datasets were downloaded from the KernScores database in MIDI format.\footnote{\url{http://kern.ccarh.org/}.} In total, the composite corpus includes the complete catalogues for Beethoven's string quartets and piano sonatas, Joplin's rags, and Chopin's piano works, and consists of over 1,000 compositions containing more than 1 million chord tokens.

\begin{table}[t!]
	\centering
	\begin{threeparttable}
		\begin{tabular}{l l S[table-format=4.0] S[table-format=7.0] S[table-format=4.0]}
			\toprule
			\textit{Composer} & \textit{Genre} & \textit{N}$_{\text{pieces}}$ & {\textit{N}$_{\text{tokens}}$} & {\textit{N}$_{\text{types}}$} \\
			\midrule
			Bach & Chorale & 370 & 35237 & 786\\
			Haydn & Quartet & 210 & 159579 & 1472\\
			Mozart & Quartet & 82 & 78201 & 1289\\
			Beethoven & Quartet & 70* & 132896 & 1699\\
			Mozart & Piano & 51 & 92279 & 833\\
			Beethoven & Piano & 102* & 176370 & 1332\\
			Chopin & Piano & 155* & 147827 & 1790\\
			Joplin & Piano & 47* & 43848 & 854\\
			Assorted & Symphony & 29 & 147549 & 2420\\
			\addlinespace[.2cm]
			\multicolumn{2}{r}{\textit{Total}} & 1116 & 1013786 & 2590\\
			\bottomrule
		\end{tabular}%
		\begin{tablenotes}
			\footnotesize
			\item \textit{Note}. * denotes the complete catalogue.
		\end{tablenotes}
	\end{threeparttable}
	\caption{Datasets and descriptive statistics for the corpus.}
	\vspace{-.3cm}
	\label{tab:corpus}%
\end{table}%

\subsection{Chord Representation Scheme}\label{sec:representationscheme}

\begin{figure}[t!]
	\hspace{.1em}
	\def\svgwidth{\columnwidth}
	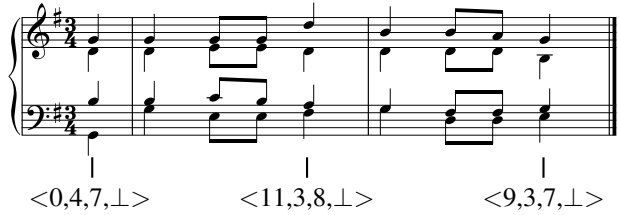
	
	\vspace{-.1cm}
	
	\caption{Full expansion of Bach, ``Aus meines Herzens Grunde'', mm. 1--2. Three chord onsets are shown with the tonal encoding scheme described in Section \ref{sec:representationscheme} for illustrative purposes.}
	\vspace{-.3cm}
	\label{fig:bach_example2}
\end{figure}

To derive chord progressions from symbolic corpora using data-driven methods, music analysis software frameworks typically perform a \textit{full expansion} of the symbolic encoding, which duplicates overlapping note events at every unique onset time. Shown in \figref{fig:bach_example2}, expansion identifies 9 unique onset times in the first two measures of Bach's chorale harmonization, ``Aus meines Herzens Grunde.''

Previous studies have represented each chord according to the simultaneous relations between its note-event members (e.g., vertical intervals) \cite{Sears:2016}, the sequential relations between its chord-event neighbors (e.g., melodic intervals) \cite{Conklin:2002}, or some combination of the two \cite{Quinn:2010}. For the purposes of this study, we have adopted a chord typology that models every possible combination of note events in the corpus. The encoding scheme consists of an ordered tuple ($S, I$) for each chord onset in the sequence, where $S$ is a set of up to three intervals above the bass in semitones modulo the octave (i.e., 12), resulting in $13^3$ (or 2197) possible combinations;\footnote{The value of each vertical interval is either undefined (denoted by $\perp$), or represents one of twelve possible interval classes, where 0 denotes a perfect unison or octave, 7 denotes a perfect fifth, and so on.} and $I$ is the chromatic scale degree (again modulo the octave) of the bass, where 0 represents the tonic, 7 the dominant, and so on.

Because this encoding scheme makes no distinction between chord tones and non-chord tones, the syntactic domain of chord types is still very large. To reduce the domain to a more reasonable number, we have excluded pitch class repetitions in $S$ (i.e., voice doublings), and we have allowed permutations. Following \cite{Quinn:2010}, the assumption here is that the precise location and repeated appearance of a given interval are inconsequential to the identity of the chord. By allowing permutations, the major triads $\langle4, 7, 0\rangle$ and $\langle7, 4, 0\rangle$ therefore reduce to $\langle4, 7, \perp\rangle$. Similarly, by eliminating repetitions, the chords $\langle4, 4, 10\rangle$ and $\langle4, 10, 10\rangle$ reduce to $\langle4, 10, \perp\rangle$. This procedure restricts the domain to $233$ unique chord types in $S$ (i.e., when $I$ is undefined).

To determine the underlying tonal context of each chord onset, we employ the key-finding algorithm in \cite{Albrecht:2013}, which tends to outperform other distributional methods (with an accuracy of around 90\% for both major and minor keys). Since the movements in this dataset typically feature modulations, we compute the Pearson correlation between the distributional weights in the selected key-finding algorithm and the pitch-class distribution identified in a moving window of 16 quarter-note beats and centered around each chord onset in the sequence. The algorithm interprets the passage in Figure \ref{fig:bach_example2} in G major, for example, so the bass note of the first harmony is 0 (i.e., the tonic).

\section{Language Models}\label{sec:languagemodels}

The goal of language models is to estimate the probability of event $e_i$ given a preceding sequence of events $e_1$ to $e_{i-1}$, notated here as $e^{i-1}_1$. In principle, these models predict $e_i$ by acquiring knowledge through unsupervised statistical learning of a training corpus, with the model architecture determining how this learning process takes place. For this study we examine the two most common and best-performing language models in the NLP community: (1) Markovian finite-context (or $n$-gram) models using the PPM algorithm, and (2) recurrent neural networks (RNNs) using both long short-term memory (LSTM) layers and gated recurrent units (GRUs).

\subsection{Finite Context Models}

Context models estimate the probability of each event in a sequence by stipulating a global order bound (or deterministic context) such that $p(e_i)$ depends only on the previous $n-1$ events, or $p(e_i|e^{i-1}_{(i-n)+1})$. For this reason, context models are also sometimes called $n$-gram models, since the sequence $e^{i}_{(i-n)+1}$ is an \textit{n}-gram consisting of a context $e^{i-1}_{(i-n)+1}$, and a single-event prediction $e_i$. These models first acquire the frequency counts for a collection of sequences from a training set, and then apply these counts to estimate the probability distribution governing the identity of $e_i$ in a test sequence using maximum likelihood (ML) estimation.

Unfortunately, the number of potential $n$-grams decreases dramatically as the value of $n$ increases, so high-order models often suffer from the \textit{zero-frequency problem}, in which $n$-grams encountered in the test set do not appear in the training set \cite{Witten:1991}. The most common solution to this problem has been the \textit{Prediction by Partial Match} (PPM) algorithm, which adjusts the ML estimate for $e_i$ by combining (or \textit{smoothing}) predictions generated at higher orders with less sparsely estimated predictions from lower orders \cite{Cleary:1984}. Specifically, PPM assigns some portion of the probability mass to accommodate predictions that do not appear in the training set using an \textit{escape method}. The best-performing smoothing method is called \textit{mixtures} (or \textit{interpolated smoothing}), which computes a weighted combination of higher order and lower order models for every event in the sequence.

\subsubsection{Model Selection}

To implement this model architecture, we apply the variable-order Markov model (called \textit{IDyOM}) developed in \cite{Pearce:2005}.\footnote{The model is available for download: \url{http://code.soundsoftware.ac.uk/projects/idyom-project}} The model accommodates many possible configurations based on the selected global order bound, escape method, and training type. Rather than select a global order bound, researchers typically prefer an extension to PPM called PPM*, which uses simple heuristics to determine the optimal high-order context length for $e_i$, and which has been shown to outperform the traditional PPM scheme in several prediction tasks (e.g., \cite{Pearce:2004}), so we apply that extension here. Regarding the escape method, recent studies have demonstrated the potential of \textit{method C} to minimize model uncertainty in melodic and harmonic prediction tasks \cite{Hedges:2016,Pearce:2004}, so we also employ that method here.

To improve model performance, Finite Context models often separately estimate and then combine two subordinate models trained on differed subsets of the corpus: a \textit{long-term} model (LTM+), which is trained on the entire corpus; and a \textit{short-term} (or \textit{cache}) model (STM), which is initially empty for each individual composition and then is trained incrementally (e.g., \cite{Conklin:1995}). As a result, the LTM+ reflects inter-opus statistics from a large corpus of compositions, whereas the STM only reflects intra-opus statistics, some of which may be specific to that composition. Finally, the model implemented here also includes a model that combines the LTM+ and STM models using a weighted geometric mean (BOTH+) \cite{Pearce:2005b}. Thus, we report the LTM+, STM, and BOTH+ models for the analyses that follow.\footnote{The models featuring the + symbol represent both the statistics from the training set \textit{and} the statistics from that portion of the test set that has already been predicted.}

\subsection{Recurrent Neural Networks}

\newcommand{\y}{\bm{y}}
\newcommand{\x}{\bm{x}}
\renewcommand{\o}{\bm{o}}
\newcommand{\e}{\bm{e}}
\newcommand{\h}{\bm{h}}
\renewcommand{\W}{\bm{W}\!}
\newcommand{\Yb}{\bm{Y}}
\newcommand{\Y}{\mathcal{Y}}
\renewcommand{\L}{\mathcal{L}}

Recurrent Neural Networks (RNNs) are powerful models designed for sequential modelling tasks.
RNNs transform an input sequence $\x_1^N$ to an output sequence $\o_1^N$
through a non-linear projection into a hidden layer $\h_1^N$, parameterised by
weight matrices $\W_{hx}$, $\W_{hh}$ and $\W_{oh}$:
\begin{align}
  \h_i &= \sigma_h\left(\W_{hx}\x_i + \W_{hh}\h_{i-1}\right) \label{eq:hidden_layer}\\
  \o_i &= \sigma_o\left(\W_{oh}\h_i\right) \label{eq:output_layer},
\end{align}
where $\sigma_h$ and $\sigma_o$ are the activation functions for the hidden layer
(e.g.\ the sigmoid function), and the output layer (e.g.\ the softmax),
respectively. We excluded bias terms for simplicity.

\begin{figure}[t!]
	\hspace{.1em}
	\def\svgwidth{\columnwidth}
	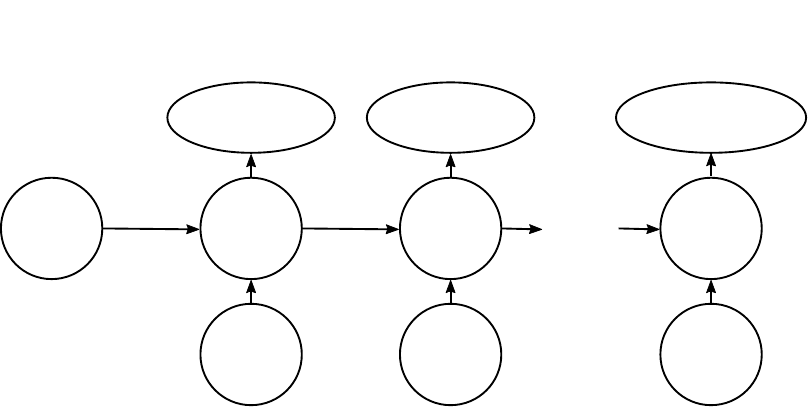
	
	\vspace{-.1cm}
	
	\caption{The basic architecture for an RNN-based language model. This model can easily accommodate more recurrent hidden layers or include additional skip-connections between the input and each hidden layer or the output. The first input, $e_0$, is a dummy symbol without an associated chord.}
	\vspace{-.3cm}
	\label{fig:rnn_language_model}
\end{figure}

RNNs have become popular models for natural language processing
due to their superior performance
compared to Finite Context models \cite{Mikolov2010}. Here, the input at each time step $i$ is a
(learnable) vector representation of the preceding symbol, $\bm{v}(e_{i-1})$. The network's output
$\o_i \in \mathbb{R}^{N_{\text{types}}}$ is interpreted as the conditional probability
over the next symbol, $p\left(e_i \mid e_1^{i-1}\right)$. As outlined in
Figure~\ref{fig:rnn_language_model}, this probability depends on \emph{all}
preceding symbols through the recurrent connection in the hidden layer.

During training, the categorical cross-entropy between the output $\o_i$ and
the true chord symbol is minimised by adapting the weight matrices in
Eqs.~\ref{eq:hidden_layer} and~\ref{eq:output_layer} using stochastic gradient
descent and back-propagation through time. However, this training procedure
suffers from vanishing and exploding gradients because of the recursive dot product in Eq.~\ref{eq:hidden_layer}. The latter problem can be averted by clipping the gradient values; the former, however, is trickier to prevent, and necessitates more complex recurrent structures such as the long short-term memory unit (LSTM)~\cite{Hochreiter1997} or the gated recurrent unit
(GRU)~\cite{Cho2014}. These units have become standard features of RNN-based language modeling architectures~\cite{Melis2018}.

\subsubsection{Model Selection}

Selecting good hyper-parameters is crucial for neural networks to perform well.
To this end, we performed a number of preliminary experiments to tune the
networks. Our final architecture comprises two layers of 128 recurrent units
each (either LSTM or GRU), a learnable input embedding of 64 dimensions
(i.e.\, $\bm{v}(\cdot)$ maps each chord class to a vector in
$\mathbb{R}^{64}$), and skip connections between the input and all other
layers.

RNNs are prone to over-fit the training data. We use the network's performance
on held-out data to identify this issue. Since we employ 4-fold
cross-validation (see Sec.~\ref{sec:experiments} for details), we hold out one
of the three training folds as a validation set. If the results on these data do
not improve for 10 epochs, we stop training and select the model with the
lowest cross-entropy on the validation data.

We trained the networks for a maximum of 200 epochs, using stochastic gradient
descent with a mini-batch size of 4. Each of these 4 data points is a sequence
of at most 300 chords. The gradient updates are scaled using the Adam update
rule~\cite{Kingma2014} with standard parameters. To prevent exploding gradients,
we clip gradient values larger than 1.

\section{Experiments}\label{sec:experiments}

\subsection{Evaluation}

To evaluate performance using a more refined method than one simply based on the accuracy of the model's prediction, we use a statistic called \textit{corpus cross-entropy}, denoted by $H_m$.
\begin{equation}
H_m(p_m,e^j_1) = - \frac{1}{j}\displaystyle\sum^{j}_{i=1}\log_2p_m(e_i|e^{i-1}_1).
\end{equation}
$H_m$ represents the average information content for the model probabilities estimated by $p_m$ over all $e$ in the sequence $e_1^j$. That is, cross-entropy provides an estimate of how uncertain a model is, on average, when predicting a given \textit{sequence} of events \cite{Pearce:2004}, regardless of whether the correct symbol for each event was assigned the highest probability in the distribution. 

Finally, we employ 4-fold cross-validation stratified by dataset for both model architectures, using cross-entropy as a measure of performance.

\subsection{Results}

We first compare the average cross-entropy estimates across the entire corpus using Finite Context models and RNNs, and then examine the estimates across datasets for the best performing model configuration from each architecture. We conclude by examining the differences between these models in a regression analysis.

\subsubsection{Comparing Models}\label{sec:comparingmodels}

\begin{table}[b!]
	\centering
	\begin{threeparttable}
		\renewcommand{\TPTminimum}{\columnwidth} 
		\makebox[\columnwidth]{
			\begin{tabular}{llccc}
				\toprule
				\multicolumn{3}{l}{\textit{Model Type}} & $H_m$ & \textit{CI}\tnote{a} \\
				\midrule
				\multicolumn{5}{l}{\textit{Finite Context}} \\
				& LTM+   &       & 4.895      & 4.811--4.978  \\
				& STM   &       & 6.710      & 6.600--6.820  \\
				& BOTH+  &       & 4.893 & 4.800--4.966  \\
				\multicolumn{5}{l}{\textit{Recurrent Neural Network}} \\
				& LSTM  &       & 5.583       & 5.539--5.626 \\
				& GRU   &       & 5.600      & 5.551--5.645  \\
				& & & & \\ [-2ex]
				\bottomrule
			\end{tabular}%
		}
		\begin{tablenotes}
			\footnotesize
			\item[a] CI refers to the 95\% bootstrap confidence interval of $H_m$ using the bias-corrected and accelerated percentile method with 1000 replicates.
		\end{tablenotes}
	\end{threeparttable}
	\caption{Model comparison using cross-entropy as an evaluation metric.}
	\label{tab:modelcomparison}%
\end{table}%

\begin{figure*}[t!]
	\def\svgwidth{.9\textwidth}
	\centering
	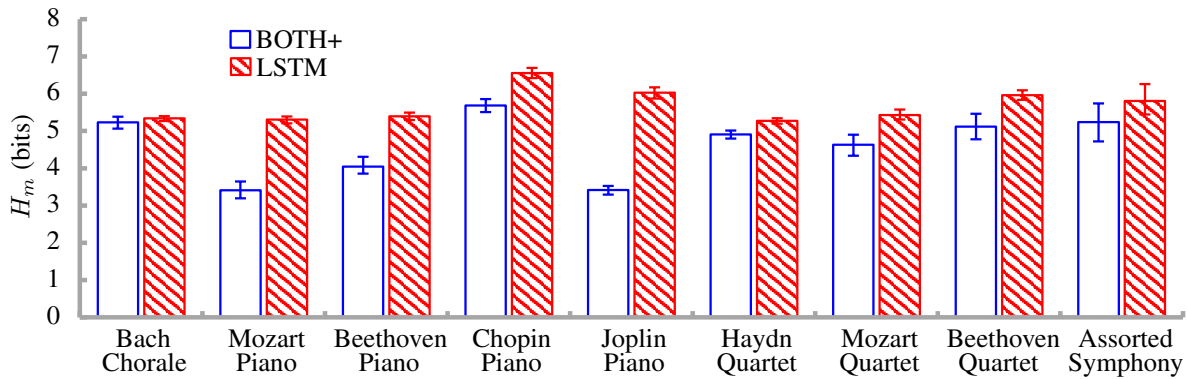
	
	\vspace{-.1cm}
	
	\caption{Bar plots of the best-performing model configurations from the Finite Context (BOTH+) and RNN (LSTM) models. Whiskers represent the 95\% bootstrap confidence interval of the mean using the bias-corrected and accelerated percentile method with 1000 replicates.}
	\vspace{-.3cm}
	\label{fig:barplot}
\end{figure*}

\tabref{tab:modelcomparison} presents the average cross-entropy estimates for each model configuration. For the purposes of statistical inference, we also include the 95\% bootstrap confidence interval using the bias-corrected and accelerated percentile method \cite{DiCiccio:1996}. For the Finite Context models, BOTH+ produced the lowest cross-entropy estimates on average, though the difference between BOTH+ and LTM+ was negligible. STM was the worst performing model overall, which is unsurprising given the restrictions placed on the model's training parameters (i.e., that it only trains on the already-predicted portion of the test set). 

Of the RNN models, LSTM slightly outperformed GRU, but again this difference was negligible. What is more, the long-term Finite Context models (BOTH+ and LTM+) significantly outperformed both RNNs. This finding could suggest that context models are better suited to music corpora, since the datasets for melodic and harmonic prediction are generally miniscule relative to those in the NLP community \cite{Korzeniowksi:2018}. The encoding scheme for this study also produced a large vocabulary (2590 symbols), so the PPM* algorithm might be useful when the model is forced to predict particularly rare types in the corpus. 


\subsubsection{Comparing Datasets}\label{sec:comparingdatasets}

To identify the differences between these models for each of the datasets in the corpus, \figref{fig:barplot} presents the bar plots for the best-performing model configurations from each model architecture: BOTH+ from the Finite Context model, and LSTM from the RNN model. On average, BOTH+ produced the lowest cross-entropy estimates for the piano datasets (Mozart, Beethoven, Joplin), but much higher estimates for the other datasets. This effect was not observed for LSTM, however, with the datasets' genre --- chorale, piano work, quartet, and symphony --- apparently playing no role in the model's overall performance.

The difference between these two model architectures for the Joplin and Mozart piano datasets is particularly striking. Given the degree to which piano works generally consist of fewer homorhythmic textures relative to the other genres in this corpus, it could be the case that the piano datasets feature a larger proportion of rare, monophonic chord types relative to the other datasets. The next section examines this hypothesis using a regression model.

\subsubsection{A Regression Model}

Given the complexity of the corpus, a number of factors might explain the performance of these models. Thus, we have included the following five predictors in a multiple linear regression (MLR) model to explain the average cross-entropy estimates for the compositions in the corpus ($N=1136$):\footnote{Four of the 1116 compositions were further subdivided in the selected datasets, producing an additional 20 sequences in the analyses: Beethoven, Quartet No. 6, Op. 18, iv (2); Chopin, Op. 12 (2); Mozart, Piano Sonata No. 6, K. 284, iii (13); Mozart, Piano Sonata No. 11, K. 331, i (7).}

\begin{description}  
	\item[$N_{\text{tokens}}$] Cache (i.e., STM) and RNN-based language models often benefit from datasets that feature longer sequences by exploiting statistical regularities in the portion of the test sequence that was already predicted. Thus, $N_{\text{tokens}}$ represents the number of tokens in each sequence. Compositions featuring more tokens should receive lower cross-entropy estimates on average.

	\item[$N_{\text{types}}$] Language models struggle with data sparsity as $n$ increases (i.e., the zero-frequency problem). One solution is to select corpora for which the vocabulary of possible distinct types is relatively small. Thus, $N_{\text{types}}$ represents the number of types in each sequence. Compositions with larger vocabularies should receive higher cross-entropy estimates on average. 
	
	\item[Improbable] Events that occur with low probability in the zeroth-order distribution are particularly difficult to predict due to the data sparsity problem just mentioned. Thus, \textit{Improbable} represents the proportion of tokens in each sequence that appear in the bottom 10\% of types in the zeroth-order probability distribution. Compositions with a large proportion of these particularly rare types should receive higher cross-entropy estimates on average.  
	
	\item[Monophonic] Chorales feature homorhythmic textures in which each temporal onset includes multiple coincident pitch events. The chord types representing these tokens should be particularly common in this corpus, but some genres might also feature polyphonic textures in which the number of coincident events is potentially quite low (e.g., piano). Thus, \textit{Monophonic} represents the proportion of tokens in each sequence that consist of only one pitch event. Compositions with a large proportion of these monophonic events should receive higher cross-entropy estimates on average.
	
	\item[Repetition] Compared to chord-class corpora, data-driven corpora are far more likely to feature adjacent repetitions of tokens. Thus, \textit{Repetition} represents the proportion of tokens in each sequence that feature adjacent repetitions. Compositions with a large proportion of repetitions should receive lower cross-entropy estimates on average. 
		
\end{description}

\tabref{tab:regression} presents the results of a stepwise regression analysis predicting the average cross-entropy estimates with the aforementioned predictors. $R^2$ refers to the fit of the model, where a value of 1 indicates that the model accounts for all of the variance in the outcome variable (i.e., a perfectly linear relationship between the predictors and the cross-entropy estimates). The slope of the line measured for each predictor, denoted by $\beta$, represents the change in the outcome resulting from a unit change in the predictor.

For the Finite Context model (BOTH+), four of the five predictors explained 53\% of the variance in the cross-entropy estimates. As predicted, cross-entropy decreased as the number of tokens increased, suggesting that the model learned from past tokens in the sequence. What is more, cross-entropy increased as the vocabulary increased, as well as when the proportion of monophonic or improbable tokens increased, though the latter two predictors had little effect on the model.

For the RNN model, the effect of these predictors was strikingly different. In this case, cross-entropy increased with the proportion of improbable events. Note that this predictor played only a minor role for the Finite Context model, which suggests PPM* may be responsible for the model's superior performance. For the remaining predictors, cross-entropy estimates decreased when the proportion of adjacent repeated tokens increased. Like the Finite Context model, the RNN model also struggled when the proportion of monophonic tokens increased, but benefited from longer sequences featuring smaller vocabularies.

\begin{table}[t!]
	\centering
	\begin{threeparttable}
		\renewcommand{\TPTminimum}{\columnwidth} 
		\makebox[\columnwidth]{
			\begin{tabular}{llS[table-format=1.3]c}
				\toprule
				\textit{Model} & \textit{Predictors} & \multicolumn{1}{c}{$\beta$} & \multicolumn{1}{c}{$R^2$} \\
				\midrule
				\multicolumn{1}{l}{BOTH+} &       &       &  \\
				& $N_{\text{tokens}}$ & -2.079 & .212 \\
				& $N_{\text{types}}$ & 1.860 & .506 \\
				& \textit{Monophoni}c  & 0.233 & .506 \\
				& \textit{Improbable} & 0.076 & .530 \\
				\multicolumn{1}{l}{LSTM} &       &       &  \\
				& \textit{Improbable} & 0.463 & .318 \\
				& \textit{Repetition} & -0.558 & .375 \\
				& $N_{\text{types}}$ & 0.817 & .504 \\
				& \textit{Monophonic} & 0.452 & .568 \\
				& $N_{\text{tokens}}$ & -0.554 & .591 \\
				\bottomrule
			\end{tabular}%
		}
		\begin{tablenotes}
			\footnotesize
			\item \textit{Note}. Each predictor appears in the order specified by stepwise selection, with $R^2$ estimated at each step. However, $\beta$ presents the standardized betas estimated in the model's final step.
		\end{tablenotes}
	\end{threeparttable}
	\caption{Stepwise regression analysis predicting the average $H_m$ estimated for each composition from the best-performing model configurations with characteristic features of the corpus.}
	\vspace{-.3cm}
	\label{tab:regression}%
\end{table}%

\section{Conclusion}\label{sec:discussion}

This study examined the potential for language models to predict chords in a large-scale corpus of tonal compositions from the common-practice period. To that end, we developed a flexible chord representation scheme that (1) made minimal a priori assumptions about the chord typology underlying tonal music, and (2) allowed us to create a much larger corpus relative to those based on chord annotations. Our findings demonstrate that Finite Context models outperform RNNs, particularly in piano datasets, which suggests PPM* is responsible for the superior performance, since it assigns a portion of the probability mass to potentially rare, as-yet-unseen types. A regression analysis generally confirmed this hypothesis, with LSTM struggling to predict the improbable types from the piano datasets. 

To our knowledge, this is the first language-modeling study to use such a large vocabulary of chord types, though this approach is far more common in the NLP community, where the selected corpus can sometimes contain millions of distinct word types. Our goal in doing so was to bridge the gulf between the most current data-driven methods for melodic and harmonic prediction on the one hand \cite{Sears:2018}, and applications of chord typologies for the creation of corpora using expert analysts on the other \cite{burgoyne_expert_2011}. Indeed, despite recent efforts to determine the efficacy of language models for annotated corpora \cite{Korzeniowksi:2018,DiGiorgi:2017}, relatively little has been done to develop unsupervised methods for the discovery of tonal harmony in predictive contexts.

One serious limitation of the architectures examined in this study is their unwavering commitment to the surface. Rather than skipping seemingly inconsequential onsets, such as those containing embellishing tones or repetitions, these models predict every onset in their path. As a result, the model configurations examined here attempted to predict tonal (pitch) content rather than tonal harmonic progressions per se. In our view, word class models could provide the necessary bridge between the bottom-up and top-down approaches just described by reducing the vocabulary of surface simultaneities to its most essential harmonies \cite{Brown:1992a}. Along with prediction tasks, these models could then be adapted for sequence generation and automatic harmonic analysis, and in so doing, provide converging evidence that the statistical regularities characterizing a tonal corpus also reflect the \textit{order} in which its constituent harmonies occur.   


\section{Acknowledgments}\label{sec:acknowledgements}

This project has received funding from the European Research Council (ERC) under the European Union's Horizon 2020 research and innovation programme (grant agreement $\text{n}^\circ$ 670035).


\bibliography{ISMIR2018,filipsrefs}

\end{document}